\documentclass[prl,aps,showpacs,groupedaddress,longbibliography,superscriptaddress,twocolumn,toc=flat,biblatex,footinbib]{revtex4-2}
\usepackage[table]{xcolor}
\usepackage[colorlinks=false]{hyperref}
\usepackage[utf8]{inputenc}
\usepackage{color}
\usepackage{bbm} 
\usepackage[english]{babel}
\usepackage{multirow}
\usepackage{amsfonts,amsmath,amssymb,stmaryrd}
\usepackage{braket}
\usepackage{graphicx}
\usepackage{subfigure}  % use for side-by-side figures
\usepackage{dsfont}
\usepackage{epsfig}
\usepackage{mathrsfs}
\usepackage{verbatim}
\usepackage{centernot}
\usepackage{longtable}
\usepackage{nicefrac}
\usepackage[framemethod=tikz]{mdframed}
\usepackage{siunitx}
\usepackage[nolist]{acronym}
\usepackage{placeins}
\usepackage{soul}
\usepackage{array}
\usepackage{cancel,ifthen}
\usepackage{bm}	% bold in math mode
\usepackage{siunitx}

\begin{document}

%\preprint{APS/123-QED}

\title{{Generation of Phonons with Angular Momentum \\
 During Ultrafast Demagnetization} 
}% 

\author{M. S. Mrudul}
\email[]{mrudul.muraleedharan@physics.uu.se}
\affiliation{Department of Physics and Astronomy, Uppsala University, P.\ O.\ Box 516, S-751 20 Uppsala, Sweden}

\author{Markus Wei{\ss}enhofer}
\email[]{markus.weissenhofer@fu-berlin.de}
 \affiliation{Department of Physics and Astronomy, Uppsala University, P.\ O.\ Box 516, S-751 20 Uppsala, Sweden}
\affiliation{Department of Physics, Freie Universit{\"a}t Berlin, Arnimallee 14, D-14195 Berlin, Germany}

\author{Peter M. Oppeneer}
\email[]{peter.oppeneer@physics.uu.se}
\affiliation{Department of Physics and Astronomy, Uppsala University, P.\ O.\ Box 516, S-751 20 Uppsala, Sweden}

\date{\today}% It is always \today, today,

\begin{abstract}
A major question in the field of femtosecond laser-induced demagnetization is whereto the angular momentum lost by the electrons is transferred. Recent ultrafast electron diffraction measurements [Tauchert \textit{et al.}, Nature {\bf 602}, 73 (2022)] suggest that this angular momentum is transferred to the rotational motion of atoms on a sub-picosecond timescale, but a theory confirmation of this proposition has yet to be given. Here we investigate the coupled electron-nuclear dynamics during ultrafast demagnetization of L1$_0$ FePt, using Ehrenfest nuclear dynamics simulations combined with the time-dependent density functional theory (TDDFT) framework. We demonstrate that atomic rotations appear, i.e., the generation of phonons carrying finite angular momentum following ultrafast demagnetization. We further show that both ultrafast demagnetization and the generation of phonons with angular momentum arise from symmetry constraints imposed by the spin-orbit coupling, thus providing insight in spin-phonon interaction at ultrafast timescales.
\end{abstract}

\maketitle
{\textit{Introduction.}}
 {The observation of ultrafast demagnetization unveiled the potential of intense laser pulses to manipulate magnetization on femtosecond timescales~\cite{beaurepaire1996ultrafast,hohlfeld1997nonequilibrium,scholl1997ultrafast}. These experiments revolutionized the field of magnetism by providing a novel, time-resolved perspective on out-of-equilibrium {spin} dynamics~\cite{kirilyuk2010ultrafast,Carva2017}. Despite its profound implications, the fundamental mechanisms responsible for the ultrafast demagnetization remain a subject of intense debate. The proposed demagnetization channels include laser-induced electronic spin-flip excitations, scattering mechanisms involving electrons, phonons, and magnons, as well as spin transport mechanisms~\cite{Zhang2000,cinchetti2006spin,carpene2008dynamics,koopmans2010explaining,battiato2010superdiffusive,schellekens2013comparing, mueller2013feedback,carpene2015ultrafast,krieger2015laser,turgut2016stoner,acharya2020ultrafast,SCHEID2022169596,weissenhofer2024ultrafast}. One of the long-standing challenges in this field is to identify the processes that transfer the angular momentum from the spin subsystem {to another subsystem \cite{kirilyuk2010ultrafast,Carva2017}}. 
 
{A century ago, Einstein and de Haas showed in a seminal work that demagnetization of a macroscopic object leads to a macroscopic volume rotation \cite{Einstein}. 
While ultrafast demagnetization may well involve different dissipation channels, a transfer of spin angular momentum to phonons has frequently been proposed to take place in distinct ways
\cite{koopmans2010explaining,Carva2013,griepe2023evidence,Essert2011,Illg2013,Garanin2015,holanda2018detecting}, but has been difficult to confirm in experiments \cite{Eich2017,turgut2016stoner}.} Recent femtosecond time-resolved X-ray diffraction~\cite{dornes2019ultrafast} and ultrafast electron diffraction~\cite{tauchert2022polarized} experiments revealed intriguing lattice dynamics following ultrafast demagnetization, suggesting that the lattice serves as the primary angular momentum sink in the process. In particular, Dornes \textit{et al}. showed that demagnetization induces unbalanced forces on the surface of ferromagnetic iron, in line with an ultrafast Einstein-de Haas effect~\cite{dornes2019ultrafast}. Tauchert \textit{et al}.\ showed that time-resolved electron diffraction from ferromagnetic nickel, following demagnetization, exhibits an anisotropic pattern sensitive to the initial magnetization direction~\cite{tauchert2022polarized}. Comparing their experimental results with a microscopic model, they found strong indications that the lost angular momentum is initially transferred to atomic rotational motion before converting into macroscopic lattice dynamics.

Such atomic rotations are coherent populations of phonons carrying angular momentum. For doubly degenerate phonon modes, for which the eigenmodes correspond to atomic vibrations along two mutually orthogonal directions, a linear superposition of these degenerate modes also forms a valid eigenmode. One can hence construct an alternative basis of right- and left-circularly polarized eigenmodes, each carrying opposite angular momentum~\cite{zhang2014angular}. 
Recent works have revealed intricate connections between the polarization state of phonon modes and magnetic properties of a material~\cite{ren2021phonon,shin2018phonon,holanda2018detecting,cui2023chirality,luo2023large,davies2024phononic,juraschek2019orbital,weissenhofer2024truly}. It is in addition well known that ultrafast, intense laser excitation can generate {coherent} phonons, even when the pulse duration is significantly shorter than the phonon oscillation period~\cite{de1985femtosecond, ishioka2006coherent, cho1990subpicosecond, melnikov2003coherent, shinohara2010coherent, hase2005ultrafast,Henighan2016}. However, a key question is whether ultrafast demagnetization can be shown to induce coherent excitation of phonons carrying angular momentum.

In this Letter, we perform state-of-the-art \textit{ab initio} calculations of laser-induced ultrafast demagnetization in the L1$_0$ ordered phase of FePt. By coupling electronic and nuclear dynamics, we demonstrate the generation of phonons with finite angular momentum following ultrafast demagnetization. We further highlight the crucial role of spin-orbit coupling (SOC) in both the demagnetization process and the generation of phonons with angular momentum. Our findings are schematically summarized in Fig.~\ref{fig:schematic}: 
excitation with a femtosecond laser pulse leads to an ultrafast reduction of the atomic magnetic moments and initiates the phase-coherent rotational motion of the Fe and Pt atoms.

\begin{figure}
    \centering
 \includegraphics[width=0.9\linewidth]{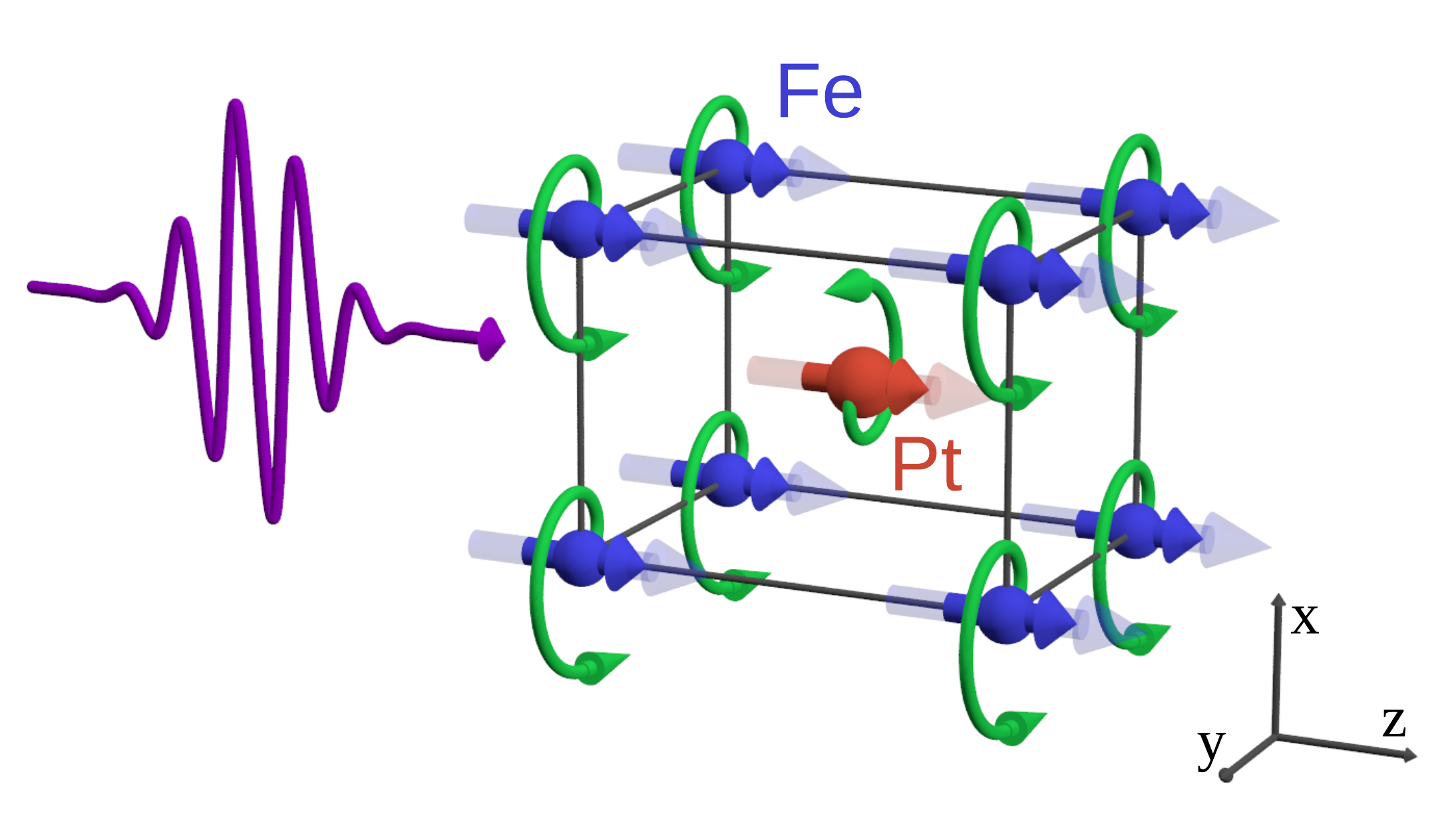}
 \vspace*{-0.4cm}
    \caption{Graphical illustration of the generation of phonons with angular momentum following ultrafast demagnetization.  The blue and red spheres represent Fe and Pt atoms,
    {respectively}. Transparent (solid) arrows represent their magnetic moments before (after) the laser pulse (purple curve). The calculated rotational motions in nuclear coordinates (green arrows) are exaggerated for better visualization.}
    \label{fig:schematic}
\end{figure}

\textit{Methodology.} {We employ time-dependent density functional theory (TDDFT) \cite{Ullrich2011, marques2012fundamentals} to investigate the dynamics of spinful electrons under the influence of an external laser field. The time-evolution of the electronic system is described by the time-dependent Kohn-Sham equation (TDKS). For the Kohn-Sham Bloch wavefunction of the $n^{\text{th}}$ band at \textbf{k} in the Brillouin zone, $\psi_{n\bm{k}}(\textbf{r})$, the TDKS equation is given by:
\begin{equation}  
\begin{split}  
i\hbar\frac{\partial}{\partial t}\psi_{n\bm{k}}(\textbf{r},t) = \Bigl\{\frac{1}{2m}&\left(-i\hbar\nabla+\frac{e}{c}\bm{A}(t)\right)^2\hat{\sigma}_0 \\  
&+\hat{V}_{\rm KS}(\textbf{r},t) \Bigr\}\psi_{n\bm{k}}(\textbf{r},t).  
\end{split}\label{eq:TDKS}  
\end{equation}  
Here, the wavefunction $\psi_{n\bm{k}}(\textbf{r})$ is a Pauli spinor that accounts for both spin components, represented as $\psi_{n\bm{k}}(\textbf{r}) = \left[\chi_{n\textbf{k}}^\uparrow(\textbf{r}),\chi_{n\textbf{k}}^\downarrow(\textbf{r})\right]^{\rm T}$. The laser field is described by the spatially homogeneous vector potential $\bm{A}(t)$, which is related to the laser electric field $\bm{E}(t)$ by $\bm{E}= -\frac{1}{c}\frac{\partial}{\partial t}\bm{A}$, where $c$ is the speed of light. The term $\hat{\sigma}_0$ is the 2$\times$2 identity matrix. The Kohn-Sham potential, $\hat{V}_{\rm KS}(\textbf{r},t)$, consists of the Hartree potential, exchange-correlation potential (computed within the local spin density approximation \cite{perdew1981self}), electron-nuclei interaction potential, and relativistic corrections (including spin-orbit coupling). We utilize the relativistic version of the norm-conserving Hartwigsen-Goedecker-Hutter (HGH) pseudopotentials \cite{hartwigsen1998relativistic}. The time-dependent changes in $\hat{V}_{\rm KS}(\textbf{r},t)$ are treated within the adiabatic approximation of TDDFT.

\begin{figure}[t!]
    \centering
    \includegraphics[width=1\linewidth]{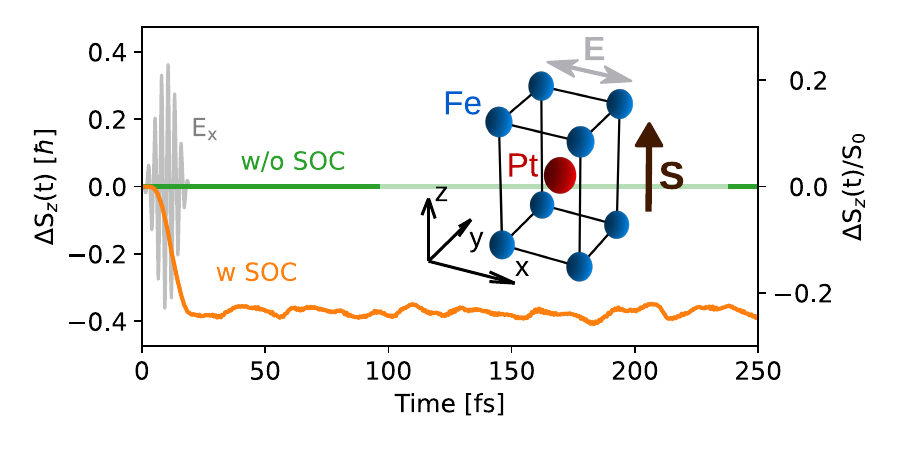}
    \vspace*{-0.6cm}
    \caption{Laser-driven dynamics of spin-angular momentum in L1$_0$ FePt, simulated with (orange) and without (green) spin-orbit coupling. The right ordinate is normalized to the equilibrium value ($S_0=S_z(0)$). The gray curve represents the temporal profile of the laser electric field. The inset illustrates the simulation setup, where the spin is aligned along the $z$ axis and the laser electric field is $x$ polarized.}
    \label{fig:demag}
\end{figure}

We incorporate Ehrenfest nuclear dynamics within the TDDFT framework to model excited phonons \cite{andrade2009modified}. In this approach, the nuclear coordinates, $\{\textbf{R}_i(t)\}$, evolve dynamically under the influence of electronic Hellmann-Feynman forces. Simultaneously, the updated nuclear configuration is used consistently to calculate $\hat{V}_{\rm KS}$ and solve the TDKS equations. The equilibrium nuclear configuration is chosen as the initial condition. For the present work,  a time step of 2.4 attoseconds was used to solve the TDKS equations as well as to update the nuclear coordinates. All calculations are performed using the \texttt{Octopus} code \cite{tancogne2020octopus}.

The Ehrenfest-TDDFT simulations are performed in the primitive tetragonal unit cell of L1$_0$ FePt, with lattice parameters $a = b =$ \SI{2.72}{\angstrom}, and $c = $ \SI{3.76}{\angstrom} \cite{maldonado2017theory}. The unit cell is sampled with a 20$\times$20$\times$30 real-space grid, and the Brillouin zone is sampled with a $6^3$ Monkhorst-Pack grid. The magnetic ground state is ferromagnetic, with magnetic moments of 2.89~$\mu_{\rm B}$ for Fe and 0.30~$\mu_{\rm B}$ for Pt along the $c$-axis \cite{maldonado2017theory,oppeneer1998magneto}. We consider a laser pulse linearly polarized along the $x$-axis with a central wavelength of \SI{800}{\nano\meter}, a pulse duration of \SI{20}{\femto\second}, and a peak intensity of \SI{2}{\tera\watt\per\square\centi\meter}. The envelope of the vector potential is modeled using a $\sin^2$ function. The inset of Fig.~\ref{fig:demag} depicts the configuration of the simulation.

{\textit{Results.}} We begin by analyzing the dynamics of spin angular momentum, calculated as  
\begin{equation}
    \textbf{S}(t) =  \frac{1}{N_\textbf{k}} \sum_{n,\textbf{k}} f_{n\textbf{k}} \left\langle \psi_{n\textbf{k}}(t)\right|\hat{\textbf{S}}\left|\psi_{n\textbf{k}}(t) \right\rangle,
\end{equation}
where $N_\textbf{k}$ is the number of sampled $k$-points, $f_{n\textbf{k}}$ represent the occupation numbers, and $\hat{\textbf{S}}$ is the spin-operator. The  operator $\hat{\textbf{S}}$, expressed in units of $\hbar$, is defined as, $\hat{\textbf{S}} = \hat{\boldsymbol{\sigma}}/2$, where $\hat{\boldsymbol{\sigma}}$ is the vector of Pauli matrices. Note that the spin magnetization can be calculated as $\textbf{M} = -(g_{\text{S}}\mu_{\text{B}}/\hbar) \textbf{S}$, where $\mu_{\rm B}$ is the Bohr magneton and $g_{\text{S}}$ is the spin g-factor.} 

Figure \ref{fig:demag} demonstrates the calculated loss of spin angular momentum due to the laser-driven electronic spin-flip excitations. It is apparent from Fig.\ \ref{fig:demag} that SOC is crucial for this demagnetization mechanism~\cite{krieger2015laser}, since without 
SOC there is no change in the spin. A net spin angular momentum of \SI{0.37}{\hbar} is lost during the laser interaction, which corresponds to a demagnetization of \SI{24}{\%}. The significant demagnetization highlights that SOC-mediated electronic spin-flip excitations is a crucial demagnetization channel in L1$_0$ FePt, as recently shown \cite{mrudul2024ab}.

\begin{figure}[t!]
    \centering
    \includegraphics[width=1\linewidth]{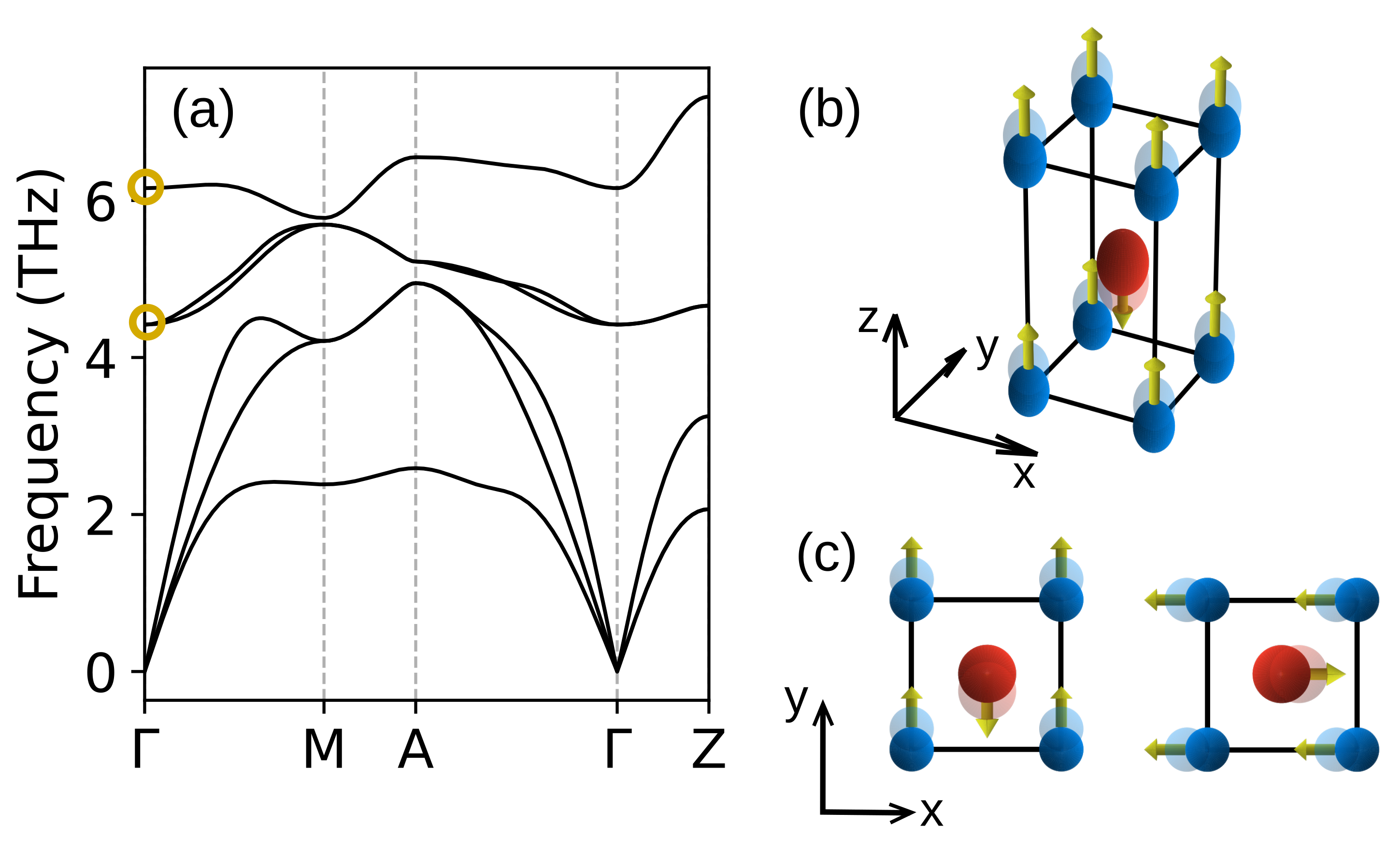}
     \caption{{(a) {\textit{Ab initio} computed} phonon dispersion of  L1$_0$ FePt with zone-center (\(\mathbf{q} = \mathbf{0}\)) optical phonon modes marked {by} yellow circles. {(b) Illustration of} zone-center optical phonon mode with frequency of \SI{6.16}{\tera\hertz}, and (c) of \SI{4.42}{\tera\hertz} (degenerate mode). The blue (red) spheres represent Fe (Pt) atoms.}}
    \label{fig:modes}
\end{figure}

\begin{figure*}[t!]
    \centering
    \includegraphics[width=1\linewidth]{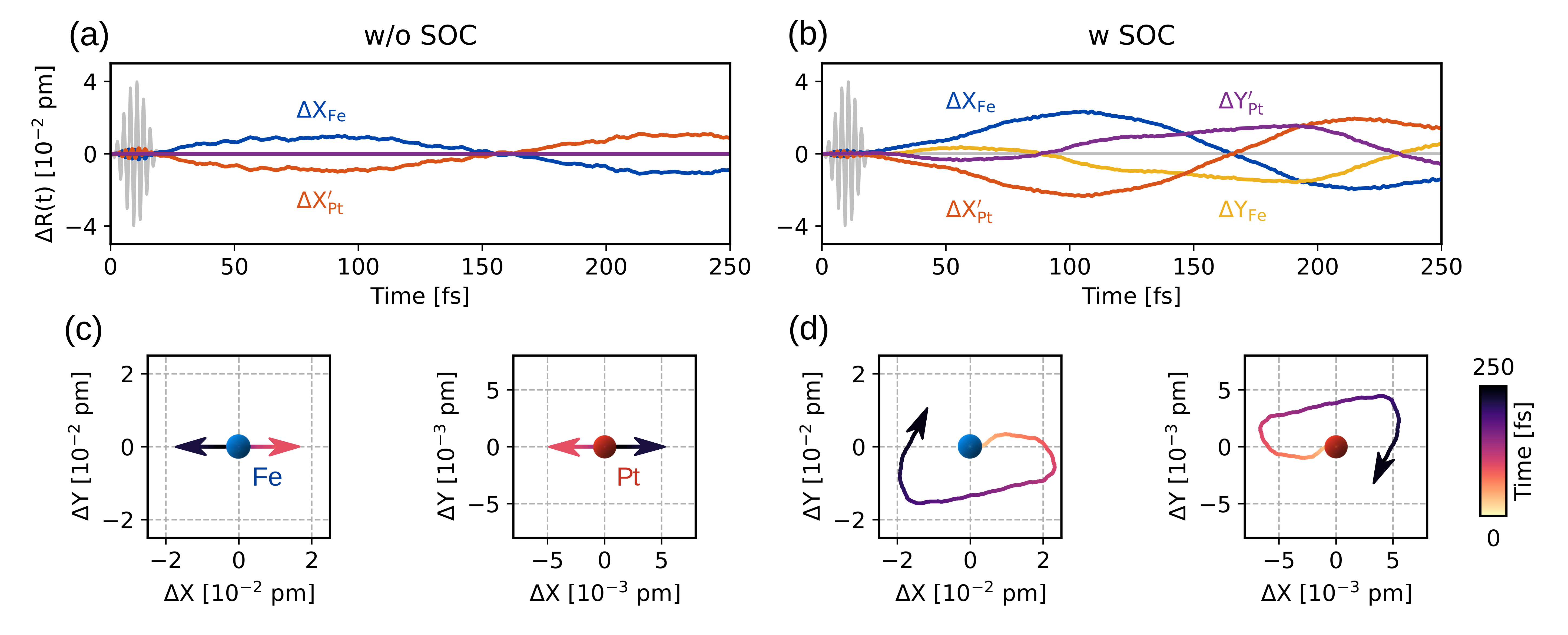}
    \vspace*{-0.5cm}
    \caption{Laser-driven nuclear dynamics following ultrafast demagnetization of L1$_0$ FePt. {Shown are the relative displacements,}  $\Delta\textbf{R}_i(t)=\textbf{R}_i(t)-\textbf{R}_i(0)$. (a) The {nuclear dynamics} calculated without SOC, and (b) with SOC. The primed coordinate of Pt is scaled as $\Delta \mathbf{R}_{\text{Pt}}^\prime = \frac{\text{M}_{\rm Pt}}{\text{M}_{\rm Fe}}  \times \Delta \mathbf{R}_{\text{Pt}}$, where $\text{M}_{\rm Fe}$ ($\text{M}_{\rm Pt}$) is the atomic mass of Fe (Pt). The gray curves in the top panels represent the temporal profile of the laser electric field. Panels (c) and (d) show the atomic trajectories of the Fe and Pt atom in the $x-y$ plane, {computed without and with SOC, respectively}.
}
    \label{fig:phonon}
\end{figure*}

Now, we discuss the ultrafast generation of phonons during ultrafast demagnetization. Our Ehrenfest-TDDFT simulations take into account the interaction of the laser-excited electrons and the zone-center (\(\mathbf{q} = \mathbf{0}\)) optical phonon modes. The phonon dispersions of FePt, highlighting the three zone-center optical modes, {are} shown in Fig.\ \ref{fig:modes}. The ground state phonon properties are calculated within the Density Functional Perturbation Theory as implemented in Quantum ESPRESSO \cite{giannozzi2009quantum, giannozzi2017advanced}. The phonon mode of frequency \SI{6.16}{\tera\hertz} corresponds to the relative motion of Fe and Pt atoms along the $z$ axis (Fig.\ \ref{fig:modes}(b)). The other two phonon modes are degenerate with a frequency of \SI{4.42}{\tera\hertz}, and are characterized by atomic vibrations along the $x$ and $y$ axis (Fig.\ \ref{fig:modes}(c)).

Figure \ref{fig:phonon} illustrates the laser-induced nuclear dynamics, where $\Delta$\textbf{R}$_i(t)=\textbf{R}_i(t)-\textbf{R}_i(0)$ represent the time-dependent displacements. 
{These are shown in Fig.\ \ref{fig:phonon}(a) and \ref{fig:phonon}(b) for calculations without and with SOC, respectively.}
It is evident from Fig.\ \ref{fig:phonon}(a) and \ref{fig:phonon}(b) that the center of mass of the cell is preserved throughout the dynamics (M$_{\rm Fe}\Delta \textbf{R}_{\rm Fe}(t)$ = $-$M$_{\rm Pt}\Delta \textbf{R}_{\rm Pt}(t)$), which is a characteristic of optical phonon modes. The generated phonons result in vibrations along the $x$ and $y$ axes, with an estimated frequency of \SI{4.17}{\tera\hertz}, in good agreement with the equilibrium properties of the degenerate phonon mode (Fig.\ \ref{fig:modes}(c)). {Although} the displacements are small, in the order of 10$^{-3}$ to 10$^{-2}$ picometers, {a clear atomic rotational motion is obtained, see Fig.\ \ref{fig:phonon}(d)}.

Interestingly, the polarization of the generated degenerate phonon modes is strongly influenced by the polarization of the laser field and SOC in the material. In the absence of SOC, the phonon mode with a vibrational axis parallel to the incident laser field is excited, resulting in linear atomic motion, see Fig.\ \ref{fig:phonon}(c). This agrees with linear forces calculated for demagnetization of Ni \cite{Zhang2022}.
{Conversely,} SOC introduces an additional transverse force on 
 atoms in an out-of-phase manner, causing them to move in an approximately elliptical trajectory with finite angular momentum, see Fig.\ \ref{fig:phonon}(d). Notably, in either case, the total linear momentum of the atomic motion vanishes ($\sum_i {\rm M}_i\dot{\textbf{R}}_i= 0$), as expected for an optical mode. In contrast, the total angular momentum arising from the rotational motion of atoms, defined as $\sum_i \Delta\textbf{R}_i \times {\rm M}_i\dot{\textbf{R}}_i$, sums up to give a finite value on the order of 10$^{-5}\hbar$.
The small magnitude of {the total angular momentum and} laser-induced atomic displacements can be attributed to the weak electron-phonon coupling of the generated optical modes in FePt \cite{maldonado2017theory}. }
{As a consequence, the here-calculated nuclear dynamics does not notably affect the spin dynamics shown in  Fig.\ \ref{fig:demag}, which we verified by comparing the results with and without Ehrenfest nuclear dynamics. 

{\textit{Discussion.}} In the following, we explain how demagnetization and the generation of phonons with angular momentum are fundamentally related to the symmetries imposed by the SOC in the Hamiltonian. The SOC potential is defined as  
\begin{equation}
    \hat{V}_{\rm SO} = \frac{1}{2m^2c^2} \hat{\textbf{S}} \cdot (\nabla \hat{V}_0 \times \hat{\textbf{p}}),
\end{equation}  
where \(\hat{V}_0\) is the electron-ion interaction potential and \(\hat{\textbf{p}}\) is the momentum operator \cite{tancogne2022effect}.  

The dynamics of the spin is governed by the commutator of the spin operator with the time-dependent electronic Hamiltonian, i.e.,  
\begin{equation}
    \frac{d{\hat{S}_z}}{dt} =  -\frac{i}{\hbar}[\hat{S}_z, \hat{\mathcal{H}}(t)] .
\end{equation}  
Since all terms in the Hamiltonian,  except for $\hat{V}_{\rm SO}$, are diagonal in spin space, their commutator with $\hat{S}_z$ vanishes \cite{Zhang2000}. This explains why the exclusive presence of $\hat{V}_{\rm SO}$ in the Hamiltonian leads to demagnetization during the laser interaction.

Now, we analyze how the transiently broken spatial symmetries affect the polarization of the generated phonons. Let us focus on the three reflection planes of L1$_0$ FePt: $\Sigma_{x}$, $\Sigma_{y}$, and $\Sigma_{z}$. For example, $\Sigma_{z}$ is defined by the invariance of the material under the transformation $(x, y, z) \xrightarrow{\Sigma_{z}} (x, y, -z)$. The excitation of $x$-polarized, $y$-polarized, and $z$-polarized optical phonon modes result in the dynamical breaking of the $\Sigma_{x}$, $\Sigma_{y}$, and $\Sigma_{z}$ reflection planes, respectively (cf.\ Fig.\ \ref{fig:modes}). This implies that the transient breaking of a reflection plane in the excited electronic density can promote the generation of the corresponding phonon mode due to the unbalanced forces exerted on ions by the electrons. In the absence of $\hat{V}_{\rm SO}$, the symmetry of the Hamiltonian is governed by the space-group symmetry of the material. Under an $x$-polarized laser field, the Hamiltonian transiently loses the reflection symmetry about the $\Sigma_{x}$ plane, i.e., $\hat{\mathcal{H}}(x,y,z,t) \neq \hat{\mathcal{H}}(-x,y,z,t)$, while preserving other reflection planes. This symmetry breaking in the Hamiltonian is subsequently manifested in the electronic density, which, via electron-ion interactions, selectively generates the $x$-polarized phonon mode.

With $\hat{V}_{\rm SO}$ included, the symmetries of the Hamiltonian are determined by both the lattice structure and the magnetic configuration, as specified by the magnetic space group \cite{dresselhaus2007group}. Consequently, new symmetry operations arise, involving coupled spatial and spin transformations. For instance, when the spin is aligned along the $z$ direction, the $\Sigma_{x}$ and $\Sigma_{y}$ reflection planes are coupled with the spin-reversal operation, while $\Sigma_z$ remains an independent symmetry \footnote{The space group and magnetic space group of L1$_0$ FePt are $P4/mmm$ and $P4/mm'm'$, respectively. The primed notation ($m \rightarrow m'$) indicates that two reflection planes are coupled with the spin-reversal operator.}. These symmetry modifications due to SOC stem from the transformation properties of angular momentum, which behaves as an axial vector. The laser-driven nuclear dynamics results from two distinct sources of transient symmetry breaking: (i) a non-relativistic effect, induced by coupling to the laser field, which breaks the $\Sigma_{x}$ reflection plane, and
(ii) a relativistic effect, that affects both $\Sigma_x$ and $\Sigma_y$ reflection planes due to coupling with the modified spin-angular momentum. These dynamically broken symmetries in the electronic system enable the generation of both $x$-polarized and $y$-polarized optical phonon modes. Since different terms in the Hamiltonian govern forces along and perpendicular to the laser field, the nuclear dynamics in the $x$ and $y$ directions differ in both amplitude and phase, resulting in an elliptical trajectory.

{Ultrafast demagnetization processes involving phonons were proposed previously
\cite{koopmans2010explaining,griepe2023evidence,Garanin2015,shin2018phonon,wu2024three,sharma2022making}.
Elliott-Yafet electron-phonon spin-flip scattering has been discussed to provide a fast channel to dissipate spin angular momentum \cite{koopmans2010explaining,Carva2013,Essert2011,griepe2023evidence}, but the accompanying ultrafast collapse of exchange splitting \cite{mueller2013feedback} has not been confirmed in experiments \cite{Eich2017,turgut2016stoner}.}
Using TDDFT calculations, Shin \textit{et al.}\ demonstrated that driving certain phonon modes in a non-magnetic two-dimensional semiconductor induces magnetization sensitive to the polarization of the driven phonon mode~\cite{shin2018phonon}. Sharma \textit{et al.}\ showed that  
specific phonon mode {displacements} in a three-dimensional ferromagnet impact demagnetization dynamics~\cite{sharma2022making}. Wu \textit{et al.}\ demonstrated that coherently generated phonons in a two-dimensional ferromagnet strongly influence the demagnetization dynamics \cite{wu2024three}. 
But none of these works showed the emergence of phonons with angular momentum upon demagnetization of the electronic system.

Generation of an atomic rotational motion was first proposed by Garanin and Chudnovsky \cite{Garanin2015}, who suggested this to be an intermediate stage before a macroscopic rotation appears as in the Einstein-de Haas effect. Our calculations confirm this prediction. Also, they are consistent with the observations of Tauchert \textit{et al.} \cite{tauchert2022polarized}, who concluded from the anisotropy of the Debeye-Waller factors that ultrafast demagnetization generates larger phonon amplitudes in the direction perpendicular to the magnetic moments. The calculated amount of phonon angular momentum at 200 fs is however quite small.
In this respect, it is important to highlight that, in magnetic materials, phonons with angular momentum can exist near high-symmetry regions throughout the Brillouin zone \cite{weissenhofer2024truly}. 
Nevertheless, it is argued that they play a crucial role in phenomena such as the phonon Hall effect \cite{strohm2005phenomenological,zhang2015chiral,chen2018chiral,park2020phonon}. 
{To describe fully the angular momentum transferred to phonons, also other phonon-phonon, electron-phonon and magnon-phonon scattering processes \cite{ren2021phonon,shin2018phonon,holanda2018detecting,cui2023chirality,luo2023large,davies2024phononic,weissenhofer2024truly} that occur across the entire Brillouin zone and at different demagnetization stages need to be considered.}
However, we have unambiguously demonstrated that the polarization properties of excited phonons during demagnetization depend on both the laser polarization and the magnetization direction.

Our calculations reveal the ultrafast generation of phonons with angular momentum, occurring within the first 100 femtoseconds of laser interaction. Thus, our work serves as an important step towards the {full}  \textit{ab initio} modeling of spin to phonon angular momentum transfer following ultrafast demagnetization \cite{tauchert2022polarized}. This work opens new avenues for theoretically exploring the polarization properties of laser-excited phonons in 
other exotic 
magnetic materials, {such as 2D ferromagnets and antiferromagnets,} offering novel insights into {as yet poorly understood} ultrafast spin-phonon interactions.

%acknowledgements
{ M.W.\ and P.M.O.\ acknowledge
 support from the German Research Foundation (Deutsche Forschungsgemeinschaft) through CRC/TRR 227 ``Ultrafast Spin Dynamics" (Project
MF, Project ID No.\ 328545488). This work has furthermore been supported by the Swedish Research Council (VR), the Knut and Alice Wallenberg Foundation (Grants No.\ 2022.0079 and No.\ 2023.0336), {and by the EIC Pathfinder OPEN grant No.\ 101129641 (OBELIX).} The calculations were enabled by resources provided by the National Academic Infrastructure for Supercomputing in Sweden (NAISS) at NSC Link\"oping partially funded by the Swedish Research Council through grant agreement No.\ 2022-06725.}

%\bibliography{main}
\providecommand{\noopsort}[1]{}\providecommand{\singleletter}[1]{#1}%

\end{document}